# Numerical studies towards practical large-eddy simulation

**J. Boudet[1], J. Caro[1], L. Shao[1], E. Lévêque[2]**

1. LMFA, Université de Lyon - Ecole Centrale de Lyon - Université Lyon 1 - INSA Lyon - UMR CNRS 5509
36 avenue Guy de Collongue, 69134 Ecully Cedex, France
2. Laboratoire de Physique, Université de Lyon - Ecole Normale Supérieure de Lyon - UMR CNRS 5672
46 allée d'Italie, 69364 Lyon Cedex 07, France

**Abstract**

Large-eddy simulation developments and validations are presented for an improved simulation of turbulent internal flows. Numerical methods are proposed according to two competing criteria: numerical qualities (precision and spectral characteristics), and adaptability to complex configurations. First, methods are tested on academic test-cases, in order to abridge with fundamental studies. Consistent results are obtained using adaptable finite volume method, with higher order advection fluxes, implicit grid filtering and "low-cost" shear-improved Smagorinsky model. This analysis particularly focuses on mean flow, fluctuations, two-point correlations and spectra. Moreover, it is shown that exponential averaging is a promising tool for LES implementation in complex geometry with deterministic unsteadiness. Finally, adaptability of the method is demonstrated by application to a configuration representative of blade-tip clearance flow in a turbomachine.

## Nomenclature

| | | | |
|---|---|---|---|
| $C_p$ | specific heat at constant pressure | $u_w$ | friction velocity |
| $C_s$ | constant of Smagorinsky | $x_i$ | coordinate in the direction $i$ |
| $c_{exp}$ | exponential average coefficient | $y^+$ | distance $y$ in wall units ($y^+ = y\, u_w / v$) |
| $c$ | airfoil chord length | $\alpha, \beta$ | filter coefficients (shape and relaxation) |
| $E_{ii}$ | energy spectrum on $u_i$ | $\Delta t$ | numerical time step |
| $e$ | internal energy | $\Delta x$ | numerical spatial step |
| $G_i$ | transfer functions | $\delta$ | channel half width |
| $k$ | turbulent kinetic energy | $\kappa$ | wavenumber |
| $l_c$ | large-scale characteristic length | $\mu$ | dynamic viscosity |
| $M$ | step of the grid turbulence experiment | $\mu_{sgs}$ | subgrid-scale viscosity |
| $P$ | static pressure | $v$ | kinematic viscosity |
| $Pr$ | Prandtl number | $\rho$ | density |
| $Pr_{sgs}$ | subgrid-scale Prandtl number | $\tau$ | time period |
| $R_{ii}$ | two-point velocity correlation | $\omega$ | non-dimensional wavenumber |
| $T$ | temperature | — | filter operator |
| $U_0$ | inflow velocity | $\sim$ | Favre operator ($\widetilde{q} = \overline{\rho q} / \overline{\rho}$) |
| $u_c$ | large-scale characteristic velocity | $< . >$ | average |
| $u_i$ | velocity component in the direction $i$ | | |

## 1. Introduction

Improvement of computational fluid dynamics for the prediction of internal flows needs progress in different but complementary directions. These are mainly:

- The enlargement of the computational extent (in space, time, towards other physics…), in order to include more physical phenomena. Schlüter et al. [1] investigate such effects in the aerodynamic computation of a complete gas-turbine.

- A better representation of turbulence, particularly for 3D interaction with complex mean field.

The present paper mainly addresses the second point. A preference is given to direct simulation of turbulence, compared to modeling of averaged effects. Indeed, increasing the complexity of turbulence modeling is of limited efficiency: only average data are computed, and validity is not guaranteed for complex flows that are not comparable to calibration cases. In comparison, direct simulation delivers more physical information (spectra, correlations...) and is more adaptable.

Large-eddy simulation (LES) [2] allows direct simulation of turbulence with a reduced number of points: only the largest turbulent eddies are represented, out of a spectral width that grows with Reynolds number. LES already represents a major tool for academic study of turbulent flows, from the transition in boundary layer [3] to separated flows on bluff bodies [4]. Application to practical flows is developing, and a particularly interesting example is given by Raverdy et al. [5] on a turbine blade. These developments also benefits from theoretical and numerical studies, such as the analysis of compressible aspects by Erlebacher et al. [6].

It can be noted here that direct representation of turbulence can also improve description of other physical aspects. For example, it provides spectra and two-point correlations used for broadband noise simulation, of major importance to study nuisance from turbo-engines. Such computations are presented on airfoils by Terracol et al. [7] or Boudet et al. [8].

The present paper intends to analyze adaptable methods for LES, from academic test-cases where turbulent physics is precisely isolated, to complex configurations and associated numerical constraints. Next section presents numerical methods, chosen for their numerical characteristics (precision, transfer function…) and their adaptability. Afterwards, the numerical methods are analyzed on two academic test-cases: isotropic turbulence and channel flow, in order to abridge with fundamental developments on LES. Finally, a section presents application to a complex configuration, representative of turbomachine blade-tip clearance flow.

## 2. Adaptable numerical methods

### General equations

The *Turb'Flow* solver uses the filtered equations of continuity, momentum and energy:

$$\frac{\partial \overline{\rho}}{\partial t} + \frac{\partial \overline{\rho}\tilde{u}_j}{\partial x_j} = 0$$

$$\frac{\partial \overline{\rho}\tilde{u}_i}{\partial t} + \frac{\partial \overline{\rho}\tilde{u}_i \tilde{u}_j}{\partial x_j} = -\frac{\partial \overline{P}}{\partial x_i} + \frac{\partial \overline{\tau}_{ij}}{\partial x_j} + \frac{\partial \Pi_{ij}}{\partial x_j}$$

$$\frac{\partial \overline{\rho}\tilde{e}_t}{\partial t} + \frac{\partial}{\partial x_j}\left[\left(\overline{\rho}\tilde{e}_t + \overline{P}\right)\tilde{u}_j\right] = \frac{\partial}{\partial x_j}\left[\tilde{u}_i\left(\overline{\tau}_{ij} + \Pi_{ij}\right)\right] + \frac{\partial}{\partial x_j}\left(\frac{\overline{\mu}C_p}{\Pr}\frac{\partial \tilde{T}}{\partial x_j}\right) + \frac{\partial \theta_j}{\partial x_j}$$

The commutation error of the filter operator with the partial derivatives is neglected. The subgrid-scale stress tensor is: $\Pi_{ij} = \overline{\rho}\tilde{u}_i\tilde{u}_j - \overline{\rho u_i u_j}$, and the subgrid-scale heat flux:

$$\theta_j = \left(\overline{\rho}\tilde{e}_t + \overline{P}\right)\tilde{u}_j - \overline{\left(\rho e_t + P\right)u_j} - \tilde{u}_i\left(\overline{\tau}_{ij} + \Pi_{ij}\right) + \overline{u_i \tau_{ij}}$$

The fluid is air, considered Newtonian:

$$\overline{\tau}_{ij} = \overline{\mu}\left(\frac{\partial \tilde{u}_i}{\partial x_j} + \frac{\partial \tilde{u}_j}{\partial x_i} - \frac{2}{3}\frac{\partial \tilde{u}_k}{\partial x_k}\delta_{ij}\right)$$

with constant dynamic viscosity. Finally, the ideal gas law is used: $\overline{P} = \overline{\rho}r\tilde{T}$  (with $r$=287 J/kg-K)

The equations are solved to obtain the filtered quantities. The possible use of a discrete filter is presented below, but the approach relying only on the natural filter of the grid is shown more reliable.

**Discretisation**

Discretisation uses finite volume approach, with explicit time marching and 5-step Runge-Kutta. For convection fluxes, value of conservative quantity $q$ is given at index $j$-1/2 by:

$$q_{j-1/2} = \frac{-q_{j-2} + 7q_{j-1} + 7q_j - q_{j+1}}{12}$$

corresponding to 4th order when computing the spatial derivative at index $j$ with: $(q_{j+1/2} - q_{j-1/2})/\Delta x$, on uniform grid. The Fourier analysis of the scheme is briefly presented here, to evaluate the capabilities regarding the representation of the turbulent spectrum. Following Lele [9], the Fourier component $q(x_j) = \exp(i\omega x_j/\Delta x)$, is considered for $\omega \in [0, \pi]$. The analytical derivative is: $q'(x_j) = (i\omega/\Delta x)\exp(i\omega x_j/\Delta x)$, while the discrete derivative is: $G_1(\omega) \times q'(x_j)$, with:

$$G_1(\omega) = \frac{8\sin(\omega) - \sin(2\omega)}{6\omega}$$

This numerical damping coefficient is presented in Fig.1. Sometimes, $G_1(\omega) \times \omega$ is presented as a modified wavenumber, but this can be misleading because the actual wavenumber (in the exponential) is not altered by the linear discrete scheme. In Fig.1, it can be noted that a wide band of the spectrum is correctly described ($G_1(\omega) \approx 1$), using logarithmic scale for physical reason. The damping becomes significant ($G_1(\omega) < 0.9$) for $\omega > 0.44\pi$, corresponding to scales smaller than 4 points per wavelength.

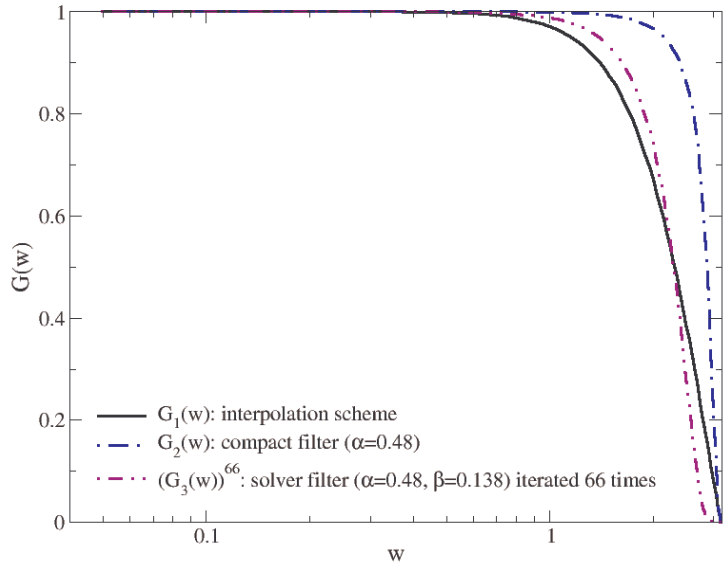

**Fig. 1** Transfer functions of the numerical schemes.

Diffusion fluxes are calculated with a second order centered scheme.

**Subgrid-scale models**

Models are required for $\Pi_{ij}$ and $\theta_j$ in order to close the system. These quantities represent the influence of the subgrid-scales on the resolved field, regarding momentum and heat flux respectively. Following the theory of Kolmogorov (cf. Pope [10]), the smallest scales are considered as mainly dissipative, at the end of the energy cascade coming from the largest eddies. This leads to the following models:

$$\Pi_{ij} = \mu_{sgs} \left( \frac{\partial \widetilde{u}_i}{\partial x_j} + \frac{\partial \widetilde{u}_j}{\partial x_i} - \frac{2}{3}\frac{\partial \widetilde{u}_k}{\partial x_k}\delta_{ij} \right)$$

$$\theta_j = \frac{\mu_{sgs} C_p}{\mathrm{Pr}_{sgs}} \frac{\partial \widetilde{T}}{\partial x_j}$$

In the present study, two models for $\mu_{sgs}$ are investigated: the standard model of Smagorinsky [11], and a shear-improved version. In both cases, $\mathrm{Pr}_{sgs} = 0.9$ and filtering is implicit (by the grid). A third approach will be presented, using an additional explicit filter to avoid accumulation of the smallest resolved scales, and $\mu_{sgs} = 0$.

● Smagorinsky model
In the well-known Smagorinsky model [11], subgrid-scale viscosity is computed locally:

$$\mu_{sgs} = \bar{\rho} C_s^2 \Delta x^2 \left| \widetilde{S} \right|$$

with: $\left| \widetilde{S} \right| = \sqrt{2 \widetilde{S}_{ij} \widetilde{S}_{ij}}$ , and $\widetilde{S}_{ij} = \frac{1}{2} \left( \frac{\partial \widetilde{u}_i}{\partial x_j} + \frac{\partial \widetilde{u}_j}{\partial x_i} \right)$

where $\Delta x$ is calculated as the cubic root of the cell volume.
The model of Smagorinsky is constructed from the theory of isotropic turbulence, where the constant can be calibrated to $C_s \approx 0.18$. Near walls, mean gradients artificially increase the value of $\mu_{sgs}$. For a plane channel flow, Deardorff [12] reduced $C_s$ to 0.1 to compensate this effect. However, it is necessary to consider more general procedures to tackle walls in complex geometries. Two recent approaches are considered here, chosen for their simplicity and their adaptability to various configurations: a shear-improved Smagorinsky model, and explicit filtering.

● Shear-improved Smagorinsky model
Lévêque et al. [13] propose the following improvement to the Smagorinsky model:

$$\mu_{sgs} = \bar{\rho} C_s^2 \Delta x^2 \left( \left| \widetilde{S} \right| - \left| < \widetilde{S} > \right| \right)$$

where the influence of the mean gradients is removed in a way that respects the energy budget for both isotropic turbulence and shear turbulence (cf. Toschi et al. [14]). In the present implementation, $\mu_{sgs}$ is forced to zero if the formula gives a negative value.
The only difficulty lies in the averaging. Spatial averaging is not adapted to complex geometries, and deterministic unsteadiness must not be removed by temporal averaging. In many practical applications (e.g. turbo-machines), an intermediate time scale $\tau$ can be defined between the turbulent time scale and the deterministic time scale. Temporal averaging will be done over this period. The implementation of a running arithmetical average over $\tau$ would require the storage of all the flow fields over this period, in order to remove the oldest samples as computation progresses. To counter this problem, exponential averaging is tested here. For a quantity $q^n$, where superscript $n$ represents temporal index:

$$< q >^n = \left( 1 - c_{\exp} \right) < q >^{n-1} + c_{\exp} q^n$$

Only storage of the average is required in this case. $c_{\exp} \in [0,1]$ is chosen according to the period $\tau$ the average has to cover. Practically, the combined influence of the data older than $\tau$ will be lower than 5% if $\left( 1 - c_{\exp} \right)^{\tau/\Delta t} = 0.05$, or equivalently:

$$c_{\exp} = 1 - 0.05^{\Delta t / \tau} \qquad (1)$$

● Explicit filtering
Bogey and Bailly [15] propose to use an additional discrete filter to continuously remove the smallest computed scales that are fed by the turbulent cascade. The filter plays the role of the subgrid-scales, and there is no other model to implement: $\Pi_{ij} = 0$ and $\theta_j = 0$. Mathew et al. [16] discuss the physical behavior reproduced by the filter as regularization. The first advantage of this approach lies in its simplicity and its generality. There is no complex physical quantity to evaluate, and no calibration. The choice of the filter parameters will be presented below. Moreover, Bogey and Bailly [17] showed that subgrid-scale viscosity can modify the effective Reynolds number of jets, and this effect is suppressed by explicit filtering.

In order to maintain the adaptability of the solver, the stencil of the filter is here limited to 5 points, using a 4th order compact filter of Lele [9]. For a conservative variable q, the filtered variable $\overline{q}^c$ is:

$$\alpha\overline{q}^c_{j-1} + \overline{q}^c_j + \alpha\overline{q}^c_{j+1} = \frac{6\alpha+5}{8}q_j + \frac{2\alpha+1}{4}\left(q_{j-1} + q_{j+1}\right) + \frac{2\alpha-1}{16}\left(q_{j-2} + q_{j+2}\right)$$

where $\alpha$ is a shape parameter. Solution of the resulting tridiagonal system is obtained by Thomas algorithm. Following again the analysis of Lele [9], the transfer function of the filter is:

$$\forall\omega\in[0,\pi]: \quad G_2(\omega) = \frac{(6\alpha+5)+4(2\alpha+1)\cos(\omega)+(2\alpha-1)\cos(2\omega)}{8(1+2\alpha\cos(\omega))}$$

and it is plotted in Fig.1 for $\alpha$=0.48. This value of $\alpha$ concentrates the damping on the highest wavenumbers, while keeping: $\forall\omega\in[0,\pi]: 0 \le G_2(\omega) \le 1$.

In the solver, the compact filter is applied to the conservative variables $q$, in the three directions of the grid sequentially, giving $\overline{q}^c$ at each time step. The actual filtered variables $\overline{q}$ are then obtained with a relaxation coefficient $\beta\in[0,1]$ : $\overline{q} = (1-\beta)q + \beta\overline{q}^c$

The overall transfer function is:

$$\forall\omega\in[0,\pi]: \quad G_3(\omega) = (1-\beta) + \beta G_2(\omega)$$

where $\beta$ allows controlling the repeated application of the filter without excessive damping at low frequencies ($G_2(\omega)$ is not exactly 1 at low $\omega$).

The value of $\beta$ has to be set according to the energy to dissipate. The idea is to filter over a band comparable to the damping band of the spatial scheme (here $G_1(\omega) < 0.9$ for $\omega > 0.44\pi$), during a time period characterizing dissipation. The Kolmogorov time scale $\tau_K$ is estimated by $\tau_K = \sqrt{\nu l_c / u_c^3}$ (cf. Pope [10]), where $l_c$ and $u_c$ are characteristic length and velocity for large-scale turbulence. Over $\tau_K$, resolved turbulence can be considered frozen and the global transfer function of the filter can be approximated by: $\omega \to (G_3(\omega))^{\tau_K / \Delta t}$. Consequently, $\beta$ is chosen to obtain: $(G_3(\omega_{c.o.}))^{\tau_K / \Delta t} = 0.5$, where $\omega_{c.o.}$ is given by: $G_1(\omega_{c.o.}) = 0.5$. This yields:

$$\beta = \frac{0.5^{\Delta t/\tau} - 1}{G_2(\omega_{c.o.}) - 1} \quad \text{where: } \tau_K = \sqrt{\nu l_c / u_c^3} \qquad (2)$$

For illustration, the transfer function $\omega \to (G_3(\omega))^{\tau_K / \Delta t}$ is plotted in Fig.1 for $\beta$=0.138 and $\tau_K / \Delta t = 66$ (values corresponding to the isotropic turbulence test-case below).

### 3. Evaluation on academic test-cases

The adaptable approaches will be first evaluated on classical academic test-cases: isotropic turbulence and bi-periodic channel flow. The present solver is dedicated to complex configurations, but these tests will abridge with more fundamental LES studies, and isolate the major physical mechanisms.

**Isotropic turbulence**

Study first focuses on the free decay of isotropic turbulence, which was particularly addressed by Comte-Bellot and Corrsin [18]. Experimental data from Kang et al. [19] are used here, because they are obtained at higher Reynolds number, and are consequently more representative of LES applications. In the experiment, turbulence is generated by a grid with step $M$=0.152m, in a flow with mean velocity $U_0$=11.25m/s. Initial measurement position is $20M$ after the grid, and final position is 48M, corresponding to free decay during time $28M/U_0$ according to Taylor hypothesis.

3D computational mesh is made of 128 points in each direction, with step $\Delta x$=0.04 m. Time step is $\Delta t$=10$^{-4}$s. Boundary faces of the cube are treated as periodic. Mean field is still air, at normal atmospheric conditions. Fluctuating velocities are initialized in the corresponding wavenumber domain, with amplitudes respecting the initial energy spectrum of the experiment, and random phases. The flow field is then transposed in the physical domain using 3D Fast Fourier Transform. Following

Kang et al. [19], computation is run for time $10M/U_0$, in order to initialize the turbulent cascade. Amplitude of the fluctuating velocity spectrum is then re-scaled on the experiment initial spectrum (cf. Fig.2). Finally, computation is run during $28M/U_0$. Results obtained with the different subgrid-scale models are presented in Fig.2.

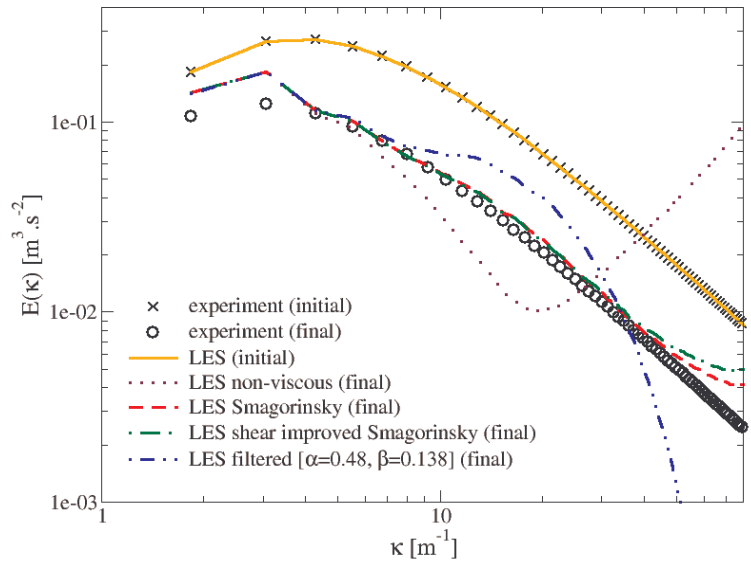

**Fig. 2** Isotropic turbulence: energy spectrum.

The models are tested in their general form, without simplification for the present academic case. Characteristic velocity is $u_c = \sqrt{2/3\,k}$, where $k$=2.38 m²/s² is initial turbulent kinetic energy, and characteristic length is $l_c$=128$\Delta x$. For the shear-improved Smagorinsky model, the time used for the averaging is the characteristic time of the large scales: $\tau = l_c / u_c \approx 4.065\text{s}$. Using Eq.(1), the coefficient of the exponential average is: $c_{exp}$=7.37 $10^{-5}$. The initialization of the exponential average field is done with still air, at normal atmospheric conditions. For explicit filtering, Eq.(2) gives the relaxation coefficient: $\beta$=0.138 (with $\tau_K$=6.57 $10^{-3}$s).

This configuration allows isolating the spectral characteristics of the methods, without walls. It is particularly demanding for adaptable methods because it is fully unsteady (statistical decay) and there is no production of turbulence. Consequently, numerical effects are accumulated, and there is no regeneration of the turbulent characteristics.

Fig.2 shows that LES is correctly initialized over the experimental spectrum. After decay, the following observations can be made.

● When no viscosity is used, energy accumulates at the highest wavenumbers, and spectrum presents an exponential growth, as expected from theory. This clearly shows the need for a subgrid-scale model. Moreover, this is the proof that present methods are not over-dissipative.

● The model of Smagorinsky allows good agreement with the experiment. A slight overprediction is observed at low wavenumbers, but this is probably the influence of the periodic boundary conditions. There is also some accumulation of energy at the highest wavenumbers (weak: logarithmic scale), but this can be explained by the uncertainty on $C_s$. This result shows the good performance of present adaptable methods, when using a model particularly designed for this case.

● Very similar results are obtained with the shear-improved Smagorinsky model. In this case, with perfect averaging (i.e. $|<\tilde{S}>| = 0$), this model actually reduces to the model of Smagorinsky.

Consequently, the present result shows the good behavior of the exponential averaging. At the end of the computation, in the exponential-averaged flow field, the standard deviation of the velocity components is $0.13 \times \sqrt{2/3\,k}$, where $k$ is the initial turbulent kinetic energy.

● Regarding the LES with explicit filtering, the damping band clearly appears at the upper end of the spectrum. The smallest resolved scales are removed, but a bump appears at the intermediate wavenumbers. Previous authors obtained very good results using such explicit filtering [15][16][17], and further investigations must be carried out to study the failure of the present computation. Such complementary study should consider how energy is transferred through the cut-off, taking into account that the smallest scales are removed. Indeed, in the present computation, the spectrum seems to indicate an accumulation of energy in the bump before the cut-off. The choice of the filter and the position of the cut-off must have an influence that could explain the present failure.

As a partial conclusion, isotropic turbulence test case shows the good spectral behavior of the present adaptable methods, with implicit filtering by the grid. Regarding the subgrid-scale modeling, explicit filtering appears to distort the spectrum, which needs to be further investigated. The Smagorinsky model, despite its good performance on isotropic turbulence, is not adapted to next configurations with walls. Consequently, only the shear-improved Smagorinsky model will be retained.

## Bi-periodic channel flow

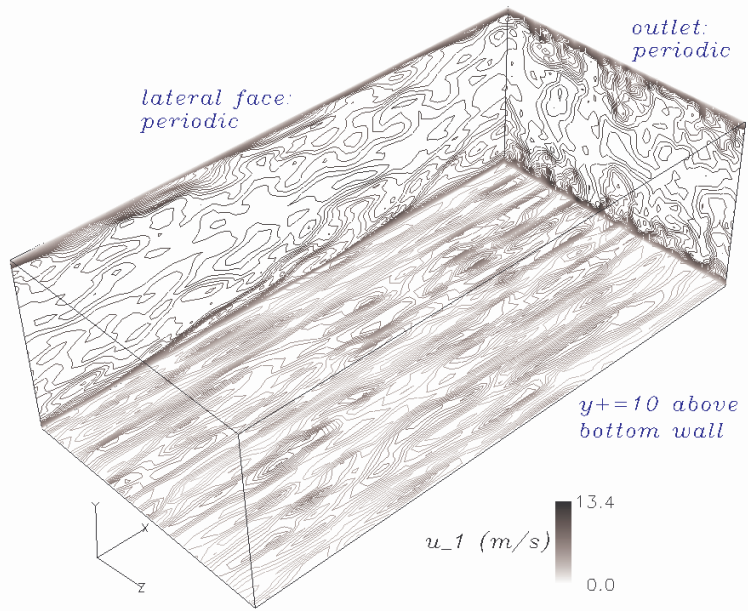

**Fig. 3** Channel: contours of $u_1$ at the final instant.

The second test-case is a bi-periodic plane channel flow at $Re_w = u_w \delta / \nu = 395$ (based on objective $u_w$ value), as studied by Moser et al. [20] with Direct Numerical Simulation (DNS). An illustration is presented in Fig.3: mean flow is oriented on $x$, walls bound the domain on $y$ direction, and periodicity is imposed on $x$ and $z$ directions. Grid extends over: $2\pi\delta \times 2\delta \times \pi\delta$ with $\delta$=0.01m. Point numbers are: 49x89x41, with tanh repartition in $y$-direction. In wall units, grid spacing is given by: $\Delta x^+$=47, $\Delta y^+$=0.5 at wall (11 points below $\Delta y^+$=10), $\Delta y^+$=22 at centerline, and $\Delta z^+$=28. Computation is initialized with Poiseuille's parabolic velocity profile for air at standard conditions (except pressure), plus 2% random velocity perturbation. Pressure is reduced to obtain a centerline Mach number of 0.2, in order to optimize convergence of the compressible solver. $x$-momentum is maintained by a source term that is adjusted dynamically to match the objective centerline velocity. Shear-improved Smagorinsky model uses exponential average with $c_{exp}$=8.837 $10^{-5}$, given $\tau = \delta/u_w$=1.69 $10^{-2}$s as characteristic time (based on objective $u_w$ value) and $\Delta t$=5 $10^{-7}$s. Post-treatment uses spatial average in $x$ and $z$ directions, plus "mirror"average in $y$ direction, and temporal average during 0.22s (90 instants).

Computation naturally transitions from perturbed Poiseuille flow to turbulent flow. This proves the present numerical methods are not over-dissipative, and the subgrid-scale model allows transition. Indeed, the shear-improved Smagorinsky is designed to yield $\mu_{sgs}$=0 in laminar flow. Resulting turbulent fluctuations are investigated below.

Fig.4 presents the mean axial velocity, compared to DNS results of Moser et al. [20]. A fairly good agreement is obtained over the different layers of the flow. The overestimate of $<u_1>/u_w$ at centerline comes from the underestimate of friction: $u_w = \sqrt{\nu \partial < u_1 > / \partial y} = 0.53$ instead of 0.59 (objective value). However, this 10% error lies within expected levels of accuracy given by Sagaut et al. [21] for quality subgrid-scale models. Thin lines aside the LES curve represent the standard deviation of velocity, taken between the exponential-averaged field at the final instant and the post-processing space-time average ($<u_1>$,0,0). A good estimate of the mean field is obtained from the exponential average, at limited constraints. Indeed, it is computed over a very limited period ($\tau = \delta/u_w = 1.69\ 10^{-2}$s), without spatial averaging. This appears as an interesting tool for LES in complex configurations, where no spatial homogeneity or stationarity can be used.

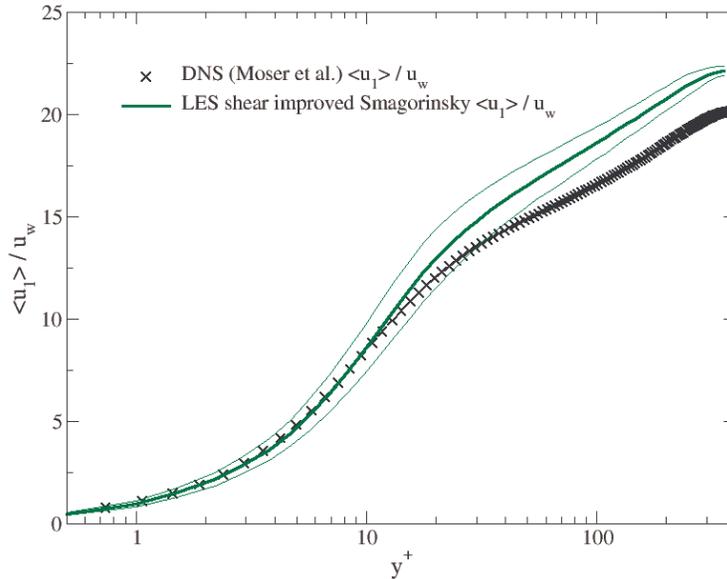

**Fig. 4** Channel: mean axial velocity (thin lines represent standard deviation of running exponential average compared to post-processing space-time average).

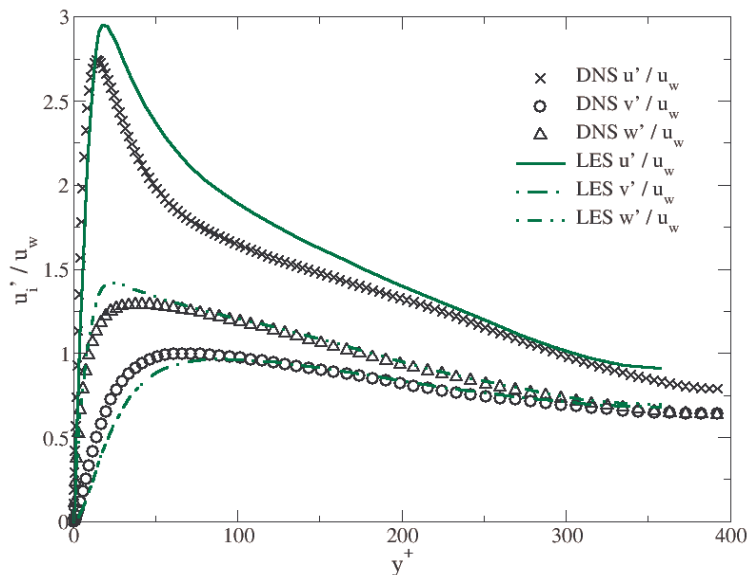

**Fig. 5** Channel: fluctuating velocity components (rms).

Fig.5 presents the fluctuating velocity components. A good agreement is obtained with DNS, concerning the levels and the evolutions. Wall turbulence anisotropy is captured with present adaptable methods.

Fig.6 presents the two-point spanwise velocity correlations:

$$R_{ii}(\Delta z) = < u_i(x, y, z).u_i(x, y, z + \Delta z) >$$

at $y^+$=10 and $y^+$=60. Post-processing average <…> covers $x$ and $z$ directions, mirror average in $y$-direction, and time. LES accurately reproduces the two-point correlations, considering the levels, the distribution between the three components of velocity, and the spanwise evolution. Comparing $y^+$=60 to $y^+$=10, the increase of the correlation length is captured, which appears through the spreading of the curves. Also, the increase of $R_{22}$ and $R_{33}$ shows some reduction of anisotropy, in agreement with DNS. This accurate representation of two-point correlations is one of the advantages of LES for practical applications. Averaged methods do not provide such information, which is particularly important for turbulence phenomena and acoustics. This prediction is achieved with a rather coarse discretisation ($\Delta z^+$=28), which results in the slope discontinuities of the LES curves in Fig.6. This is beneficial for the computational time.

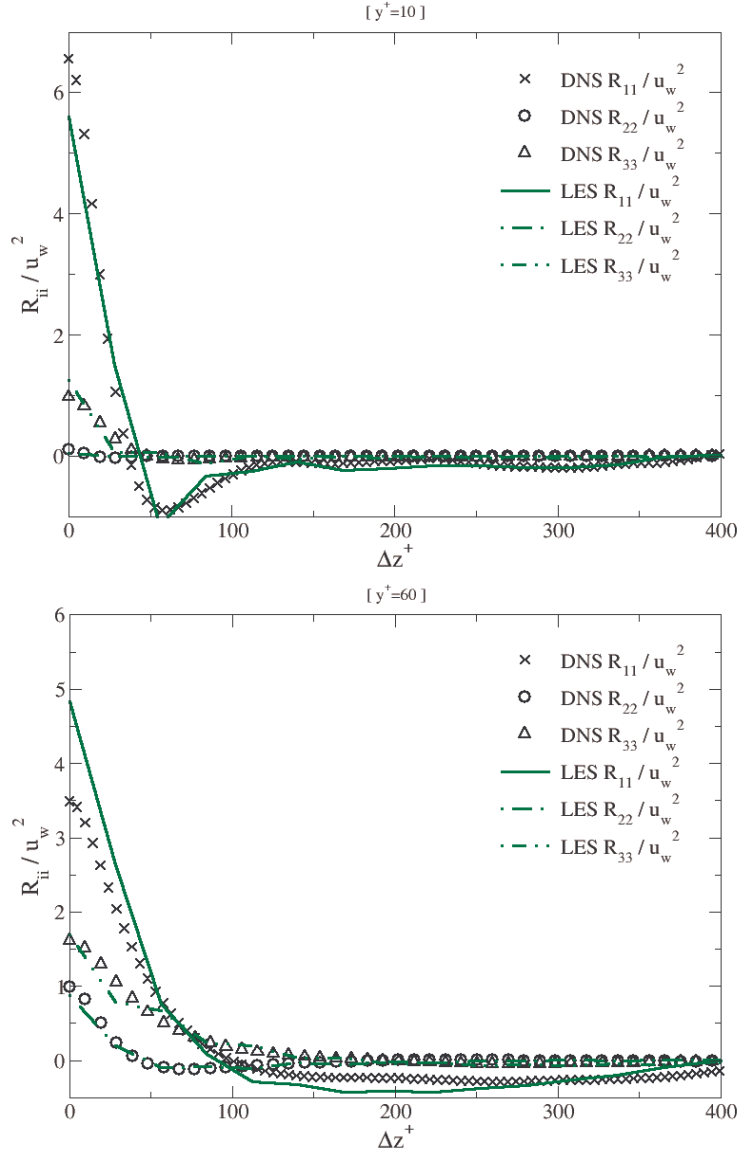

**Fig. 6** Channel: two-point spanwise correlations ($y^+$=10, and $y^+$=60).

Finally, Fig.7 presents the spanwise spectrum $E_{11}$ (energy spectrum for $u_1$) at $y^+$=10. Post-processing average covers $x$-direction, mirror average in $y$-direction, and time. This figure illustrates one of the main advantages of LES: the spectral description of the flow, for an accurate representation of turbulence eddies and acoustics. Moreover, it allows evaluating the discretisation characteristics. In comparison with reference DNS, present LES accurately represents the most energetic structures, characterized by a flat spectrum at the lower wavenumbers. Present discretisation captures the spectrum up to the decline zone, and one decade of decline is directly represented. This kind of discretisation is consistent with LES approach, which does not represent the small and poorly energetic structures. The LES curve progressively diverges from DNS at the upper end of the spectrum, probably due to the damping associated with the numerical scheme at the high wavenumbers (cf. Fig.1). The present level of physical discretisation, characterized by non-dimensional spatial steps $\Delta x^+$, $\Delta y^+$, and $\Delta z^+$ can be used as reference for future applications.

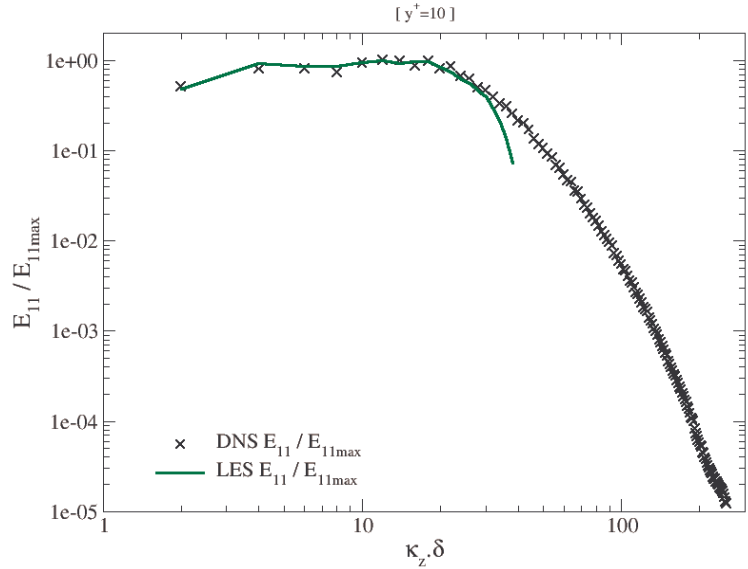

**Fig. 7** Channel: spanwise spectrum $E_{11}$ at $y^+$=10.

Academic test cases have been used to focus study on spectral dynamics and wall-turbulence. Good performances are achieved on mean flow, fluctuations, two-point correlations and spectra. Moreover, discretisation aspects have been discussed, as well as performance of the exponential averaging.

### 4. Application: tip clearance flow

This section presents a computation on a complex configuration, in order to demonstrate the adaptability of the numerical methods. The geometry is composed of a NACA5510 airfoil (chord $c$=0.2m, span: 0.2m), set perpendicularly above a flat plate, with a clearance between the plate and the airfoil tip (gap: $10^{-2}$ m). Incoming air velocity is $U_0$=70m/s. Lift on the airfoil generates a large clearance vortex, shown in Fig.8-upper. Corresponding aerodynamic and acoustic measurements are presented by Grilliat et al. [22].

This configuration is particularly representative of clearance effects in turbomachines. The objective of LES is a combined description of aerodynamics and broadband acoustics, including the turbulent physics of the detached vortical flow. The geometric complexity and the high Reynolds number ($\mathrm{Re} = U_0 c / \upsilon \approx 10^6$) require the use of adaptable numerical methods, as presented above.

This section only demonstrates the applicability of the methods, and no quantitative data are considered. Detailed numerical analysis of this configuration will be presented in a following paper. The grid is composed of about 3 million points, with near wall density $y^+ \approx 0.5$. Computation was initialized with converged RANS results on the same grid, and the clearance vortex provides natural forcing for the development of turbulent fluctuations. The shear-improved Smagorinsky model is used. For the exponential averaging, characteristic time is $\tau$=$c$/3$U_0$, corresponding to convection over one third of the chord length.

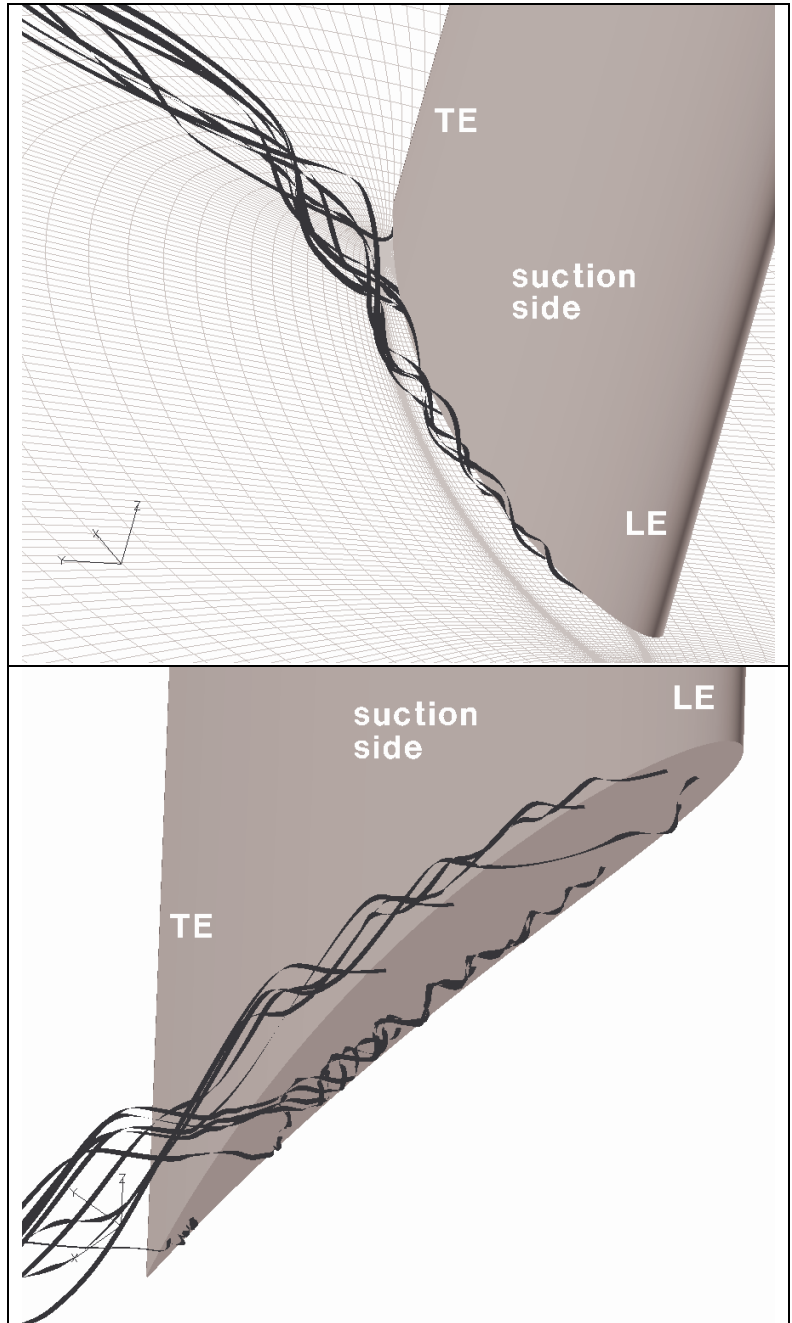

**Fig. 8** Tip clearance flow: streamlines through the gap (LE: leading edge, TE: trailing edge).

Fig.8 presents results at an instant during the development of turbulent fluctuations. In the top view, streamlines show the development of the main clearance vortex along the suction side, and its deviation by the main flow downstream of the trailing edge. In the bottom view, a secondary vortex is observed on the blade tip. It is generated by the separation of the clearance flow on the pressure side edge. LES must be particularly effective in capturing this complex vortical structure through the gap, and the associate broadband spectrum.

In summary, present adaptable numerical methods allow turbulent simulation on a complex configuration, at limited computational cost (limited discretisation stencils and number of operations). Thanks to exponential averaging, spatial homogeneity or steadiness are not required. Storage is limited to the instantaneous flow field and exponential-averaged field. Finally, the shear-improved Smagorinsky model allows the development of turbulent fluctuations.

## 5. Conclusion

The present paper focused on the development of LES for improved turbulent simulation of practical internal flows.

Numerical methods have been chosen according to two competing criteria: their performance (precision, spectral characteristics) and their adaptability to complex configurations. Also, implementation matters are discussed, such as the relationship between the filtering frequency and the spectral characteristics of the discretisation.

These methods have been tested on academic configurations, in order to abridge with previous fundamental studies on LES. Finite volume discretisation with higher order advection fluxes, implicit filtering by the grid and a "low-cost" shear-improved Smagorinsky model showed good performances on isotropic turbulence and channel flow. Mean flow, fluctuations, two-point correlations and spectra are particularly well captured. Furthermore, exponential averaging presents good performance, and appears as an interesting tool in complex configurations with no spatial homogeneity and large deterministic unsteadiness.

Finally, the adaptability of the method is confirmed by application to a complex configuration, representative of blade-tip clearance flow in turbomachine.

## Acknowledgement

This work is carried-out in the frame of Ecole Centrale de Lyon's BQR grant "Contrôle de l'écoulement de jeu en turbomachines". Dr Marc Jacob and EU consortium PROBAND are acknowledged for providing the tip-clearance test-case. Dr Lionel Gamet is acknowledged for numerical support and Dr Jean-Pierre Bertoglio for theoretical discussions on LES.